\documentstyle[aps,prl,twocolumn,psfig,epsf]{revtex}
\begin{document}
\draft
\twocolumn[
\author{C. Cabrillo, J. L. Rold{\'a}n, P. Garc{\'\i}a-Fern{\'a}ndez}
\address{Instituto de Estructura de la Materia, CSIC,
Serrano 123, 28006 Madrid, Spain.}

\date{\today}
         
\title{Quantum noise reduction in singly resonant sub/second harmonic
generation}

\maketitle
\mediumtext

\begin{abstract}
\leftskip 2.0truecm      
\rightskip -2.0truecm    
\indent
We study the quantum noise in the harmonic mode of a singly resonant
frequency doubler simultaneously driven in both modes. 
This simple extension of the frequency doubler greatly improves its 
performance as a bright squeezed light source. Specifically, for
parameters corresponding to reported experiments,
80\% of noise suppression is easily achieved, the phase of the
corresponding squeezed quadrature can be freely and easily chosen,
and the output power is nearly doubled. 
\end{abstract}

]

\narrowtext
Second Harmonic Generation has nowadays quite a long tradition as a mean of 
squeezed light generation \cite{Per88,Siz90,Kur93,Pas94,Ral95,Tsu95}. 
The preferred
experimental setup has been the doubly resonant configuration as, at least
in principle, permits arbitrarily large squeezing. However, such scheme 
has been hampered by the technical difficulties arising from keeping
the resonance in both modes simultaneously. Thus, in spite of the development
of very ingenious stabilization procedures \cite{Kur93}, 
for the moment it has been only possible to maintain the double
resonance for a few seconds. Certainly, this kind of experimental delicacy can 
hardly surprise when dealing with the generation of non-classical 
states of light. In view of such difficulties, some experimental efforts has 
been recently redirected to singly resonant configurations 
\cite{Pas94,Tsu95}. Although, the maximum noise suppression is then
limited to a 90\% \cite{Pas94}, the efforts resulted in very stable
intense squeezed light sources with degrees of squeezing
even surpassing those reported in the doubly resonant counterparts 
\cite{Tsu95}. This evolution highlights the importance of reducing
to a minimum the technical demandings of new proposals in a so 
experimentally challenging field. In this paper, we propose a
simple extension of the singly resonant scheme which, however,
can improve greatly the performance in the experimentally tested range
of parameters.

The system we want to address is such as those of \cite{Pas94} and
\cite{Tsu95}, namely, a nonlinear crystal placed inside an optical
cavity resonant with a given fundamental mode. Under phase matching
conditions, a coherent driving of the resonant mode generates and harmonic
mode that transmits freely through the cavity mirrors. Our aim here is to
study the quantum noise behavior when a coherent driving in the harmonic
mode is added. To such a task we use the two photon absorption model
developed in \cite{Coll91}. The quantum mechanical evolution equation
is given by \cite{Pas94}
\begin{eqnarray}
\label{eq:quant}
\dot {a} & = &
-\left [ \gamma + \mu a^{\dagger}a \right ]a + 
2\sqrt{\mu}\,a^{\dagger}b_{in}+ \nonumber \\
& & \sqrt{2 \gamma_{c}}\, a_{in} +
\sqrt{2\gamma_{s}}\,w_{in}\,, 
\end{eqnarray}

\noindent
where $a$ and $b_{in}$ are the annihilation operators for the fundamental
and the ingoing harmonic mode respectively, $\mu$ is proportional to
the $\chi^{(2)}$ response of the crystal, $\gamma_{c}$ is the input
coupling for the fundamental mode, $\gamma_{s}$ accounts for the
intracavity loss rate (absorption and scattering in the crystal) and 
$\gamma = \gamma_{c}+\gamma_{s}$ is the total damping rate.
Equation (\ref{eq:quant}) is complemented with the boundary conditions 
\cite{Pas94}
\begin{mathletters}
\label{eq:boundary}
\begin{eqnarray}
\label{eq:boundarya}
a_{out} & = & \sqrt{2\gamma_c}\, a - a_{in} \,,\\
\label{eq:boundaryb}
b_{out} & = & \sqrt{\mu}\,a^{2} - b_{in} \,,
\end{eqnarray}
\end{mathletters}

\noindent
from which the output spectra can be computed. Input fields are assumed
to be in coherent states. In particular, we will allow for a finite
$\langle b_{in} \rangle$ as opposed to the pure frequency doubling
configurations of \cite{Pas94} and \cite{Tsu95}. The spectra will be
computed using a standard linearization procedure around the stable
fixed points of the classical evolution equations. More explicitly,
defining fluctuation operators as
$a  = \alpha +\delta a$, 
$a_{in,out}  =  \alpha_{in,out} +\delta a_{in,out}$, and 
$b_{in,out}  =  \beta_{in,out}+\delta b_{in,out}$,
a linearization of (\ref{eq:quant}) and (\ref{eq:boundary}) yields
\begin{eqnarray}
\label{eq:linquant}
\dot{\delta a} & = &   
-\left [ \gamma + 2\mu |\alpha|^{2} 
\right]\delta a + \left [ 2\sqrt{\mu}\beta_{in} - 
\mu \alpha^{2}\right ] \delta a^{\dagger} \nonumber \\
& & +  2\sqrt{\mu}\alpha^{*}\delta b_{in}+
\sqrt{2 \gamma_{c}}\,\delta a_{in} + \sqrt{2\gamma_{s}}\,w_{in}\,,
\end{eqnarray}

\noindent
and
\begin{mathletters}
\label{eq:linboundary}
\begin{eqnarray}
\delta a_{out} & = & \sqrt{2 \gamma_{c}}\,\delta a - \delta a_{in}\,,
\label{eq:linboundarya} \\
\delta b_{out} & = & 2 \alpha \sqrt{\mu}\, \delta a - \delta b_{in}\,,
\label{eq:linboundaryb}
\end{eqnarray}
\end{mathletters}

\noindent
where $\alpha$ denotes a stationary fixed point of the classical
counterpart of Eq. (\ref{eq:quant}) and $\alpha_{in,out},\beta_{in.out}$ the 
mean values of the corresponding input and output modes.   
We will concentrate in a particular family of fixed points fulfilling 
\begin{equation}
\label{eq:fixp}
\sqrt{2\gamma_{c}} |\alpha_{in}|=
|\alpha|(\gamma + \mu n + 2 \sqrt{\mu} \,|\beta_{in}|) \,,
\end{equation}

\noindent
where $n$ denotes $|\alpha|^{2}$. The field $\alpha$ is in phase with 
$\alpha_{in}$ whose phase, $\phi$, is related with that of $\beta_{in}$,
$\varphi$, by $\varphi = 2 \phi+\pi$. The stability of the fixed point is
determined by the sign of the real part of the eigenvalues of the
corresponding drift matrix. Simple algebra yields
\begin{equation}
\label{eq:eigen}
\lambda_{\mp} = -(\gamma+2\mu n)\mp(\mu n +2 \sqrt{\mu}\,|\beta_{in}|)
\,.
\end{equation}

\noindent
An instability appears when
$2\sqrt{\mu}\,|\beta_{in}|=\gamma+\mu n$, as then $\lambda_{+}$ changes to
positive and the fixed point becomes unstable.

The squeezing phenomenon refers to a phase dependent noise such as for a
given range of phases and in a finite frequency bandwidth it becomes 
smaller than the vacuum noise.
In other words, the noise spectrum of $X_{\theta}^{out}(t)\equiv
(a(t)\exp (-i\theta)+a^{\dagger}(t) \exp(i\theta))/2$ must be below
that of a vacuum state for certain $\theta$ values. Such spectrum can
be easily minimized (maximized) on the phase yielding 
\begin{eqnarray}
\label{eq:Sw}
S(\omega)_{\mp} &  = &
\langle \delta a^{\dagger}_{out}(\omega) \delta a_{out}(-\omega)\rangle 
\mp \nonumber \\
& & \left |
\langle \delta a_{out}(\omega) \delta a_{out}(-\omega) \rangle
\right | \,,
\end{eqnarray}

\noindent
corresponding to a quadrature phase $\theta_{s} =\nu(\omega)-\pi/2$ 
($\theta_{a} =\nu(\omega)$), being $\nu(\omega)$ the phase of
$\langle \delta a_{out}(\omega) \delta a_{out}(-\omega) \rangle$. With
the normalization used the unity corresponds to the vacuum state noise. 
We will call $S_{-}(\omega)$ the squeezing spectrum and 
$S_{+}(\omega)$ the antisqueezing spectrum. They are constrained by
the Heisenberg principle so that $S_{-}(\omega)S_{+}(\omega) \geq 1$.
After tedious but simple algebra, the spectra for the 
second harmonic mode can be written as
\begin{eqnarray}
\label{eq:spectrum}
S_{\mp}(\tilde{\omega}) \; = \; 1 +  8 m |B| & & \nonumber \\
\frac{2|B|(1+2m)\mp\left[\tilde{\omega}^{2}+|B|^{2}+(1+2m)^{2}\right]}
{\left[\tilde{\omega}^{2}+(1+2m-|B|)^{2}\right]
\left [\tilde{\omega}^{2}+(1+2m+|B|)^{2}\right]}\,, & &
\end{eqnarray}

\noindent
where $\tilde{\omega} = \omega/\gamma$, $m = \mu 
|\alpha|^{2}/\gamma$, and $|B| = |\eta_{in}|+m$ with 
$ \eta_{in} \equiv 2\sqrt{\mu/\gamma} \, \beta_{in}$. With this new 
scaled variables the instability is reached at $\eta_{in}=1+m$.

The phase of the squeezed quadrature simply reduces to 
$\theta_{s} = 2 \varphi - \pi$. It is independent of the frequency or 
the photon number, something very fortunate as it allows for a control of 
the squeezing phase independent of the degree of noise reduction.
In other words, given a value of noise suppression, 
whether it is going to be amplitude squeezing, phase squeezing or 
anything in between is just a matter of choosing adequately 
$\varphi$ and $\phi$ under the constrain $\varphi = 2 \phi + \pi$.
\begin{figure}
\centerline{\psfig{figure=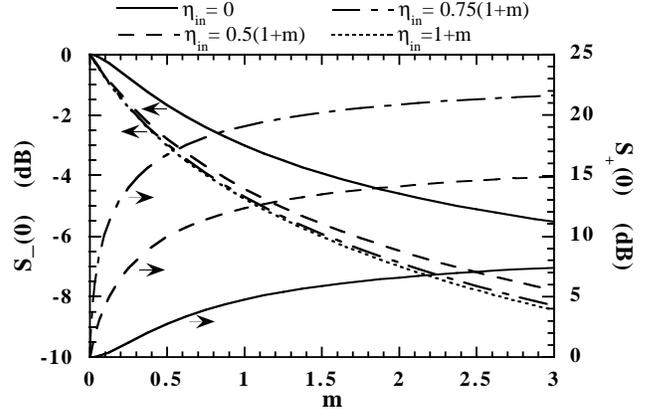,width=8.5cm,bbllx=3.2cm,bblly=11.6cm,bburx=18.0cm,bbury=22cm,clip=}}
\caption{The maximum squeezing (left) and antisqueezing (right)
in the harmonic mode in dB's with respect to the vacuum noise as a 
function of $m$.}
\label{fg:Sw}
\end{figure}

Figure \ref{fg:Sw} displays the behavior of the maximum noise reduction 
($S_{-}(\omega)$ at zero frequency) with respect to the normalized photon 
number $m$. This is a relevant physical parameter as it is proportional 
to the energy stored in the cavity. The 5.2 dB reached in \cite{Tsu95} 
corresponds approximately to $m=2.5$ (with $\eta_{in}=0$) so that 
setting the physical scale of $m$ in the reported experiments. 
Between the pure frequency doubling configuration ($\eta_{in}=0$) and 
the instability limiting case ($\eta_{in}=1+m$) two representative 
cases are shown to a 50\% and a 75\% of the instability harmonic 
driving. In spite of being at a significant ``distance'' of the  
instability, they are close to the limiting value, particularly the 75\% 
case, avoiding a wild behavior of the excess noise in the conjugate 
quadrature (the antisqueezing) also shown in the figure.  
Centering at $m=2.5$, the 
corresponding squeezing values are 5.1 dB (69\% of noise suppression) 
at $\eta_{in}=0$, 7.2 dB (80\%) at $\eta_{in}=0.5(1+m)$ and 7.6 dB
(83\%) at $\eta_{in}=0.75(1+m)$. In other words, a substantial 
squeezing enhancement is possible even at this low $m$. Obviously at 
larger values of $m$ the enhancement rapidly increases. Thus, for 
$m=20$ and at 50\% of the instability, 13.2 dB (95\%) of noise reduction 
is obtained while the excess noise remains nearly constant (16.5 dB).

\begin{figure}
\centerline{\psfig{figure=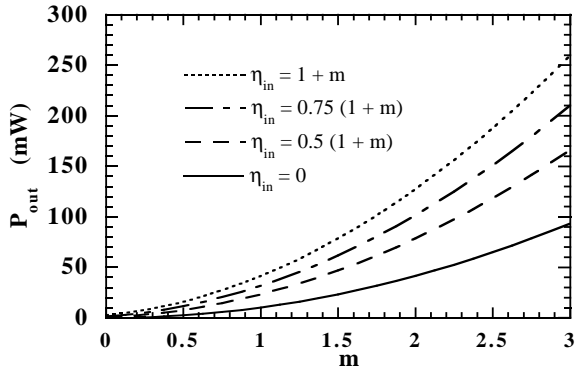,width=8.5cm,bbllx=2.5cm,bblly=11.6cm,bburx=17.3cm,bbury=20.5cm,clip=}}
\caption{The classical output power in the harmonic mode as a function 
of $m$.}
\label{fg:Pw}
\end{figure}

There is a one more advantage of the proposed system, namely, with 
the driving fields phases chosen they constructively contribute to 
the output power.  From Eq. (\ref{eq:boundaryb}) and 
after proper normalization, the classical output power results in,
\begin{equation}
\label{eq:pw}
{\rm P_{out}} \propto (2 m + \eta_{in})^{2} \,,
\end{equation} 
a direct consequence of $\alpha$ being in-phase with $\alpha_{in}$ and 
$\varphi =2\phi+\pi$. Just to fix the ideas we have taken the 
proportionality constant such as the power at $m=2.5$ is 65 mW, the 
reported value in \cite{Tsu95} at 5.2 dB of noise reduction. 
Representing ${\rm P_{out}}$ as a function of $m$ (figure \ref{fg:Pw}), 
a large enhancement appears. In particular, at $m=2.5$ the power is 
nearly doubled (118 mW) at a 50\% of the instability and more than 
doubled (151 mW) at 75\%. Quantum corrections can only raise up this 
values.

In summary, we have demonstrated that by simply adding a driving to the 
harmonic mode of the experimentally successful singly resonant 
frequency doubling scheme, a versatile and more efficient source of bright
squeezed light can be implemented for parameter values corresponding 
to the reported experiments. More specifically, the phase of the
squeezed quadrature can be controlled (independently of the 
degree of squeezing) simply adjusting the phases of the driving fields,  
80\% of squeezing can be easily reached and the output power nearly doubled.

\begin{center}
\bf Acknowledgments
\end{center}
Work supported in part by grants No. TIC95-0563-C05-03, No. PB96-00819, 
CICYT, Spain, and Comunidad de Madrid 06T/039/96.

\end{document}